\documentclass[12pt]{spieman}  
\usepackage{amsmath,amsfonts,amssymb}
\usepackage{siunitx}
\usepackage{algorithmic}
\usepackage{graphicx}
\graphicspath{{figures/}}
\DeclareGraphicsExtensions{.pdf,.eps}

\usepackage{setspace}
\usepackage{subfigure}
\usepackage{textcomp}
\usepackage{wrapfig}
\usepackage{footnote}
\usepackage{array}
\usepackage{booktabs}
\usepackage{lineno}

\title{Conversion Gain Enhancement in Standard CMOS Image Sensors}

\author[a,b,*]{Assim Boukhayma}
\affil[a]{Senbiosys, Neuchatel, Switzerland, 2000}
\affil[b]{EPFL, IMT, ICLAB, Neuchatel, Switzerland, 2000}

\begin{document} 

\maketitle

\begin{abstract}
This paper focuses on the conversion gain (CG) of pixels implementing pinned photo-diodes (PPD) and in-pixel voltage follower in standard CMOS image sensor (CIS) process. An overview of the CG expression and its impact on the noise performance of the CIS readout chain is presented. CG enhancement techniques involving process refinements and pure circuit design and pixel scheme optimization are introduced. The implementation of these techniques in a 180\,nm CIS process demonstrates a progressive enhancement of the CG by more than a factor 3 with respect to a standard reference pixel from the same foundry, allowing a better understanding of the different parasitic elements on the sense node capacitance and CG.
\end{abstract}


{\noindent \footnotesize\textbf{*}Assim Boukhayma,  \linkable{assim.boukhayma@senbiosys.com} }

\begin{spacing}{2}   

\section{Introduction}

Pinned photo-diode (PPD) based image sensors performance has been dramatically increased since their first development \cite{Tera_TED_2012} on several aspects including speed, resolution or power consumption. PPD compatibility with complementary metal oxide semiconductor (CMOS) process makes these devices an excellent candidate for a wide range of applications combining performance, miniaturization, large volume and low cost criteria \cite{Bouk_JSSC_2016, Bouk_IEEESensors_2019, Caiz_ISSCC_2019}. The sensitivity of CMOS image sensor (CIS) has been, also, remarkably improved in terms of quantum efficiency, dark current and fill-factor \cite{Tera_TED_2012}. Recently, remarkably low noise pixels, operating at room temperature, have been presented \cite{Bouk_IISW_2015, Ma_JEDS_2015, Ma:17, Seo_EDL_2015, Bouk_EDL_2020} reaching noise levels below a single electron. These improvements have been followed by demonstrations of photo-electron counting capability with CMOS image sensors without any photo-electron multiplication process  \cite{Ma_JEDS_2015,Ma:17}.

Noise reduction circuit techniques such us correlated sampling or column-level gain and bandwidth control \cite{Capp_TCAS_2020,Bouk_ICNF_2015,Yue_ISSCC_2012,Suh_Sensors_2010}, proved their efficiency in bringing the input-referred noise of CIS readout chains to sub-electron levels. Nevertheless, deep sub-electron levels close to 0.3 electrons RMS, at room temperature, could only be reached with pixel conversion gain enhancement through sense node (SN) capacitance reduction \cite{Fumi_IISW_2015,Ma_JEDS_2015,Waka_VLSI_2015,Seo_EDL_2015}. The common concept to these SN capacitance reduction techniques is the isolation of the SN from the parasitic capacitance related to the neighboring gates. This is achieved by means of process refinements or design changes that come at the cost of low pixel full well capacity (FWC) and high required voltages.

A variety of techniques involving process variations or different pixel schemes have been recently presented \cite{Fumi_IISW_2015,Ma_JEDS_2015,Waka_VLSI_2015,Seo_EDL_2015}. However, there is still some room left for demystifying the impact of the different SN area design parameters on the SN capacitance and the conversion gain in a standard CIS process. The aim of this work is to analyze the impact of the SN and in-pixel source follower (SF) design on the conversion gain and SN capacitance. The suggested way to achieve this is to implement on a same CIS array different types of pixels embedding different techniques reducing gradually the SN parasitic capacitance or increasing the conversion gain within the standard CIS process design rules boundaries. Analyzing the conversion gain measured out of the different pixel types using its formula allows extracting the contribution of the different terms and understanding the impact of each optimization step.   
  
This paper starts with an overview of the conversion gain expression and its impact on the noise performance on the CIS readout chain. Conversion gain enhancement techniques involving process refinements, pure circuit technique and pixel scheme are introduced throughout three pixel designs. The implementation of these techniques in a 180\,nm CIS process demonstrates a progressive enhancement of the conversion gain by more than a factor 3 with respect to a standard reference pixel from the same foundry. This progressive increase enables a better understanding of the impact of the SN parasitic capacitance and SF design on the conversion gain.

\section{The Conversion Gain: A Key Parameter in Low Noise PPD \& SF based CIS Pixels}

\begin{figure}[]
\centering
\includegraphics[width=0.8\linewidth]{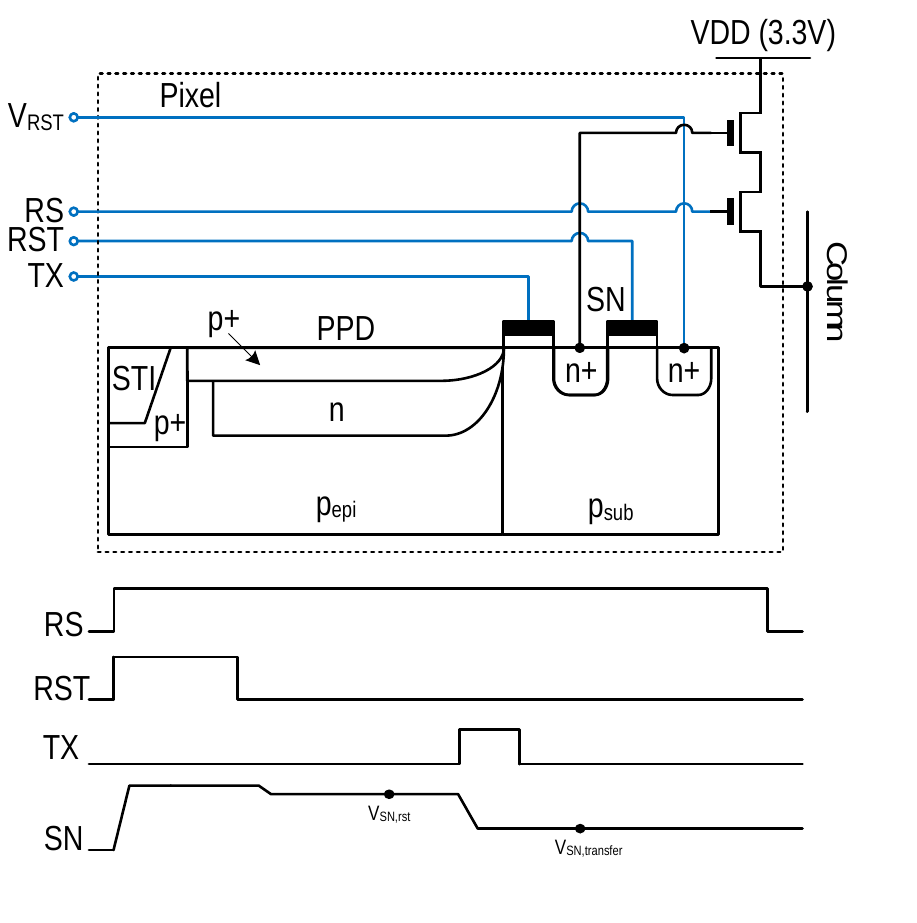}
\caption{Conventional 4T pixel schematic and corresponding readout timing diagram. The schematic depicts a cross section showing the PPD structure built on an epitaxial layer (p$_{\text{epi}}$) lightly p doped with respect to the substrate (p$_{\text{sub}}$). The borders of the PPD are protected by shallow trench isolation (STI).}\label{fig:4T_pixel}
\end{figure}

\begin{figure}[]
\centering
\includegraphics[width=0.8\linewidth]{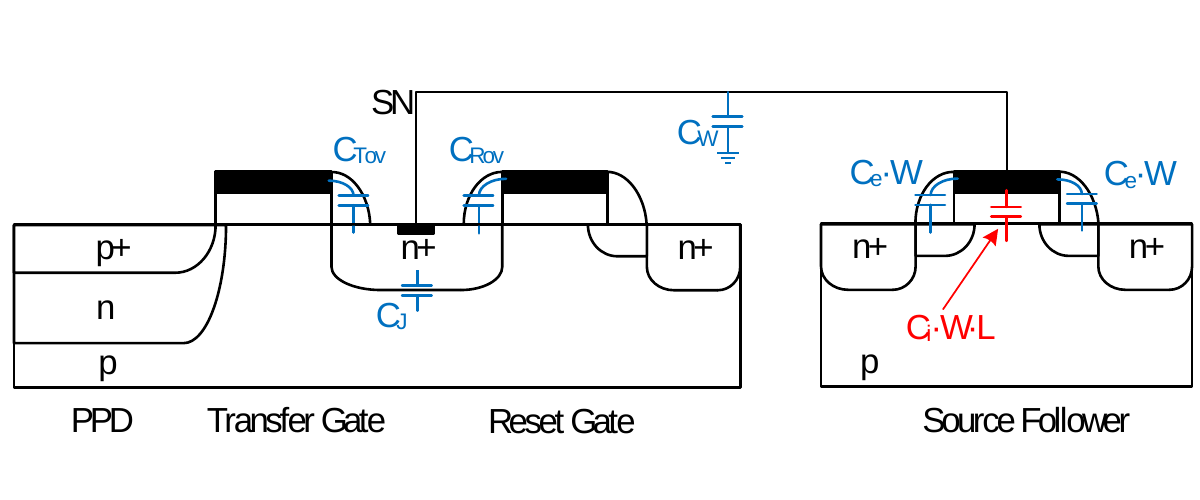}
\caption{
Cross-section schematic view of a conventional 4T pixel depicting the different parasitic elements contributing to the sense node capacitance.}\label{fig:4T_pixel_section}
\end{figure}

Fig.\,\ref{fig:4T_pixel} shows a conventional 4T pixel schematic together with its readout timing diagram. The pixel embeds a PPD integrating the photo-generated electrons, a transfer gate (TG) allowing the transfer of these integrated charges to the SN and splitting the latter from the PPD well capacitance. A reset gate (RG) allowing setting the SN to a high voltage before each transfer. When the row select (RS) switch is closed, the SF transistor buffers the voltage level of the SN to the column to be processed by the rest of the readout chain. Conventional CIS embed an array of pixels and column-level readout circuits performing a rolling readout scheme. All the pixels of a same line are readout in parallel. The column-level circuitry embeds, a correlated double sampling (CDS) scheme that takes a sample before and after the charge transfer from the PPD to the SN, an amplifier improving the signal-to-noise ratio (SNR) in case of low light conditions and an analog-to-digital converter (ADC).

The pixel conversion gain is the voltage difference that the SF creates at the column level for a single electron transferred from the PPD to the SN. Increasing this gain mitigates the impact of the noise generated at the column-level circuits which is key in low light application and also for reaching photo-electron counting capability.

The pixel SN is an area in which every fraction of a fF counts. The different elements contributing to the SN parasitic capacitance are depicted in Fig.\,\ref{fig:4T_pixel_section} and further detailed in \cite{Bouk_Sensors_2016}. The in-pixel SF is far from behaving as an ideal voltage follower. In other words, the conversion gain is not simply given by the inverse of the SN capacitance. Hence, a small signal analysis taking into account both the SN and SF parasitic capacitance is necessary to express the convesion gain. Using the small signal analysis detailed in \cite{Bouk_TED_2015,Bouk_Springer_2018}, the conversion gain, denoted  $\text{A}_{\text{CG}}$, can be formulated in the following from : 
\begin{equation}\label{eq:CG_SF}
A_{\text{CG}}=\frac{\frac{1}{n}}{C_{\text{SN}}+C_{\text{e}}W+(1-\frac{1}{n})(C_{\text{e}}W+C_{\text{i}}WL )},
\end{equation}
where, $C_{\text{e}}$ and $C_{\text{i}}$ are the SF intrinsic and extrinsic capacitance densities, $W$ and $L$ are the SF gate width and length, $C_{\text{SN}}$ is the total SN capacitance including the junction, overlap with reset and transfer gates as well as metal wires parasitic capacitances as illustrated by Fig.\ref{fig:4T_pixel_section} and $n$ is the slope factor of the source follower transistor. In saturation, the value of $n$ ranges from 1.2 to 1.6 and slowly tends to 1 for high $V_{\text{G}}$ \cite{Enz_EKV_2006}.

Eq.\,\ref{eq:CG_SF} shows that $A_{\text{CG}}$ depends on the SN capacitance, the SF parasitic capacitance and the SF body effect. 

The conversion gain plays a key role in the noise performance of the CIS readout chain. The noise generated outside the pixel is directly mitigated by a higher conversion gain since its input referred variance is simply divided by $A_{\text{CG}}^2$.
Regarding the noise generated at the level of the SF stage, both the input referred thermal and $1/f$ noise expressions share similar terms with $1/A_{\text{CG}}$. Indeed, the SF $1/f$ noise can be expressed as \cite{Bouk_TED_2015,Bouk_Springer_2018}:
\begin{equation}\label{eq:Input_Ref_1/fNoise}
\overline{Q_{1/f}^2}= \alpha_{1/f}  \frac{K_{\text{F}}}{C_{\text{ox}}^{2} W  L} (C_{\text{SN}}+2C_{\text{e}}W+ C_{\text{i}}WL)^2 ,
\end{equation}
where $K_{\text{F}}$ is a process and temperature dependent parameter, $C_{\text{ox}}$ is the oxide capacitance density and $\alpha_{1/f}$ is unite-less design dependent parameter resulting from the CDS effect on the $1/f$ noise.\\
The thermal noise  originating from the pixel SF stage can be expressed as \cite{Bouk_TED_2015,Bouk_Springer_2018}:
\begin{equation}\label{eq:Input_Ref_thermal_Noise}
\overline{Q_{\text{th}}^2}=2  f_{\text{A}}\frac{kT \gamma_{\text{SF}}}{ G_{\text{m,SF}}}(C_{\text{SN}}+2C_{\text{e}}W+ C_{\text{i}}WL)^2,
\end{equation}
where $k$ is the Boltzmann constant, $T$ the temperature, $f_{A}$ is an equivalent frequency dependent on the column-level amplifier design and bandwidth, $\gamma_{\text{SF}}$ is the SF stage excess noise factor and $G_{\text{m,SF}}$ its transconductance.

Both Eq.\,\ref{eq:Input_Ref_1/fNoise} and Eq.\,\ref{eq:Input_Ref_thermal_Noise} share a similar term having common parameters with the $A_{\text{CG}}$. Indeed, the term $C_{\text{SN}}+C_{\text{e}}W$ is common while the term $C_{\text{e}}W+C_{\text{i}}WL$ is attenuated by ($1-1/n$) in the $A_{\text{CG}}$ denominator. But in the end, reducing $C_{\text{SN}}$ and increasing $A_{\text{CG}}$ result both in a input referred noise reduction of the whole CIS readout chain especially if the $C_{\text{SN}}+C_{\text{e}}W$ dominates the $C_{\text{e}}W+C_{\text{i}}WL$ term. It is important to underline the fact that the last point is valid for SF based pixel schemes, whereas it's not valid for common source based configurations where the CG is higher thanks to the AC gain of the CS but the noise is not necessarily lower \cite{Bouk_Springer_2018,Bouk_SPIE_2014,Lott_ISSCC_2011}.

\section{CG enhancement techniques}

As shown by Eq.\,\ref{eq:CG_SF}, the CG can be enhanced by optimizing the $C_{\text{SN}}$ term on one side, and by optimizing the SF size, slope factor and parasitic capacitance on the other side. In the following, three pixel variants are suggested. The first proposes a process refinement for reducing the $C_{\text{SN}}$ term, the second combines this refinement with an SF optimization and the third variant proposes a different pixel scheme allowing further reduction of the $C_{\text{SN}}$ without any process refinements.

\subsection{Pixel variant 1}

\begin{figure}[]
\centering
\includegraphics[width=0.8\linewidth]{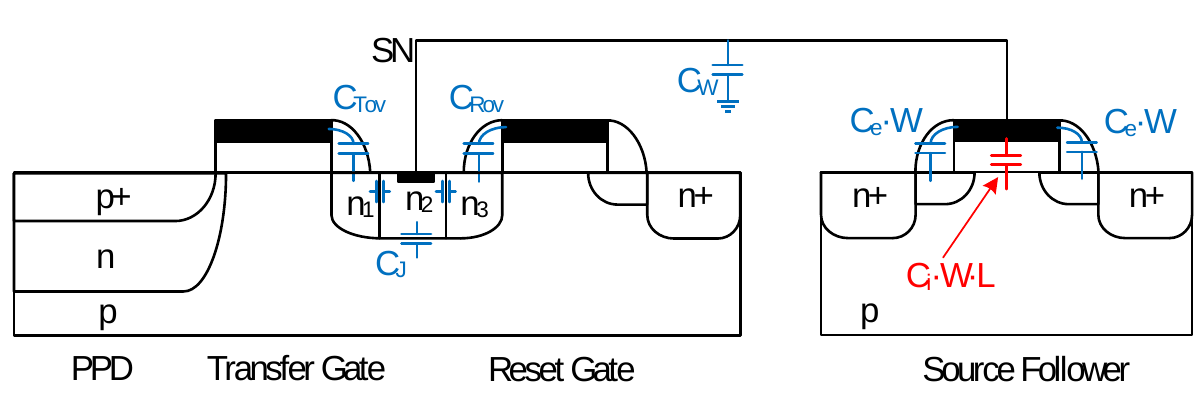}
\caption{SN doping profile improvement for reduced overlap capacitance with transfer and reset gates.}\label{fig:4T_pixel_section_new}
\end{figure}

\begin{figure}[]
\centering
\includegraphics[width=0.8\linewidth]{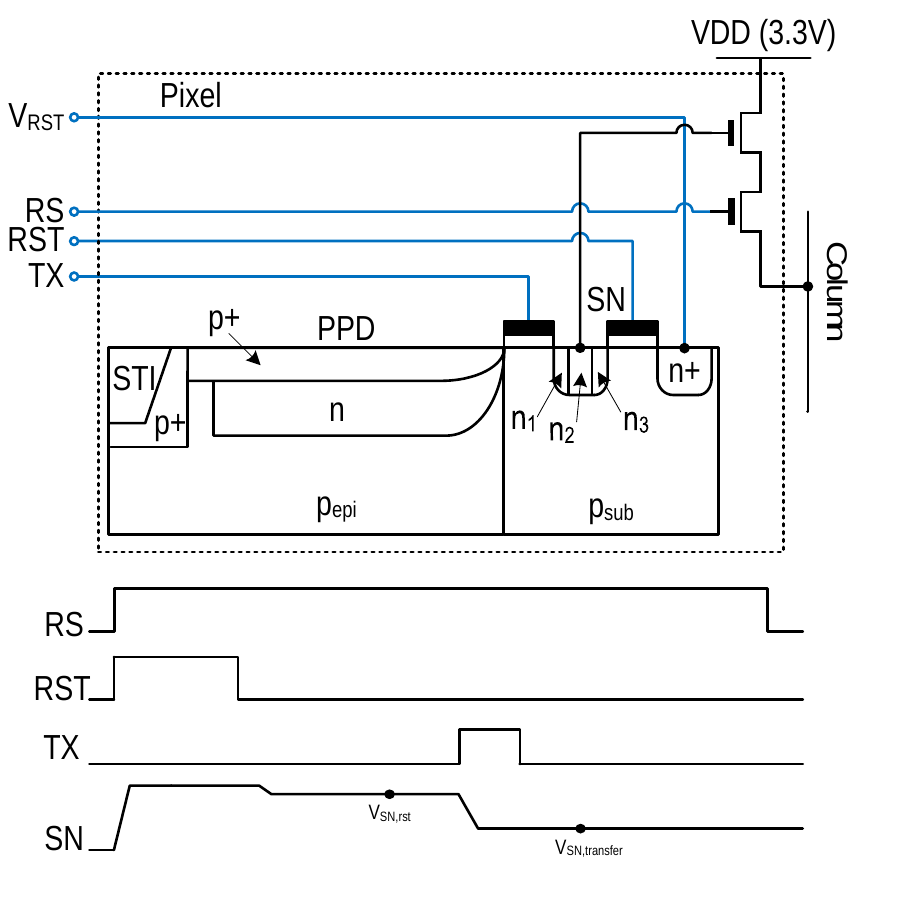}
\caption{Schematic view of the first pixel variant featuring a refined SN doping in conventional 4T pixel scheme together with the corresponding timing diagram.}\label{fig:nmos_pixel}
\end{figure}

\begin{figure}[]
\centering
\includegraphics[width=0.8\linewidth]{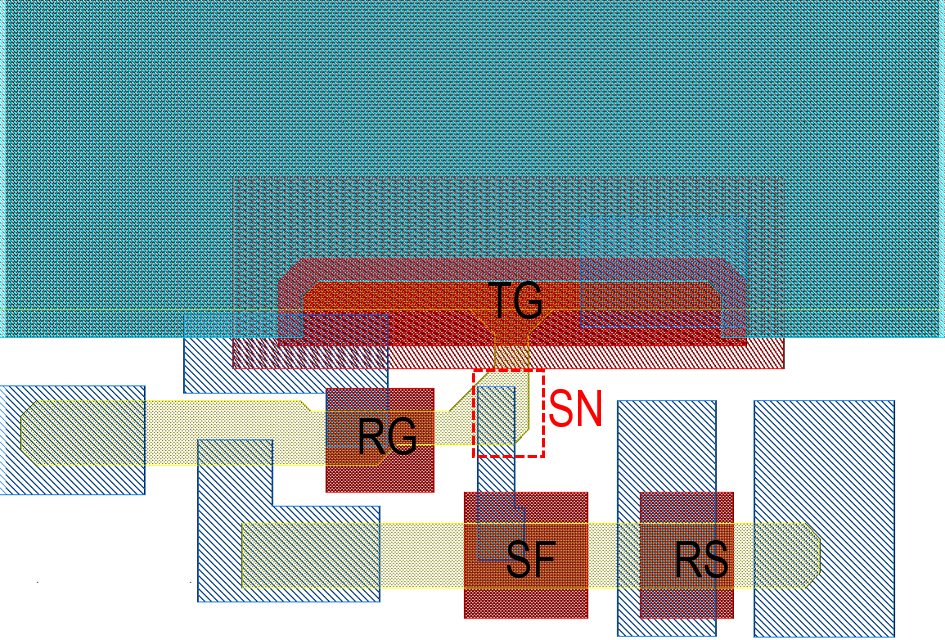}
\caption{layout view of the first pixel variant featuring a 4T scheme and implementing the SN doping optimization.}\label{fig:nmos_layout}
\end{figure}

$C_{\text{SN}}$ is the sum of the metal wiring parasitic capacitance connected to the SN, the junction capacitance of the SN and the overlapping of the SN with the transfer and reset gates. The last term dominates the $C_{\text{SN}}$ due to the large transfer gate needed for an efficient transfer and the relatively high oxide density. For instance, in the 180\,nm process used in this work, the overlap capacitance density is about 0.45\,fF/$\mu$m. This value is even prone to be higher for advanced technology nodes.
Hence, the first proposed optimization focuses on the reduction of the overlap capacitance between the SN and the transfer and reset gates. A technique similar to low doped drains (LDD) \cite{Ogur_TED_1980} is used to mitigate the overlap capacitance. Instead of uniformly doping the SN, the latter is doped with a gradually increasing concentration as shown in Fig.\,\ref{fig:4T_pixel_section_new}. The SN area overlapping with the transfer gate is doped with a concentration $n_1$ one order of magnitude lower with respect to the SN area where the metal contact is placed, $n_2$. In this way, the overlap capacitance caused by the high oxide capacitance density is mitigated by the series capacitance corresponding to the interface between the $n_1$ and $n_2$ regions. In the same way the doping concentration $n_3$ underneath the reset gate overlap with the SN area corresponds to the concentration used for LDD area in standard NMOS transistors. Such a low doping concentration together with an additional interface between $n_2$ and $n_3$ helps reducing the the overlap capacitance with the reset gate.\\
Fig.\,\ref{fig:nmos_pixel} shows the schematic of the resulting pixel and the corresponding timing diagram. This optimization has no impact on the operation scheme of the pixel. On the other hand, the layout requires additional implants in order to implement the gradual doping. Fig.\,\ref{fig:nmos_layout} shows the layout of the proposed pixel. The dashed square in the SN area marks the position of the additional implant used in the contact area. The pixel features a pitch of 12\,$\mu$m and a fill factor of 76\%.

\begin{figure}[]
\centering
\includegraphics[width=0.8\linewidth]{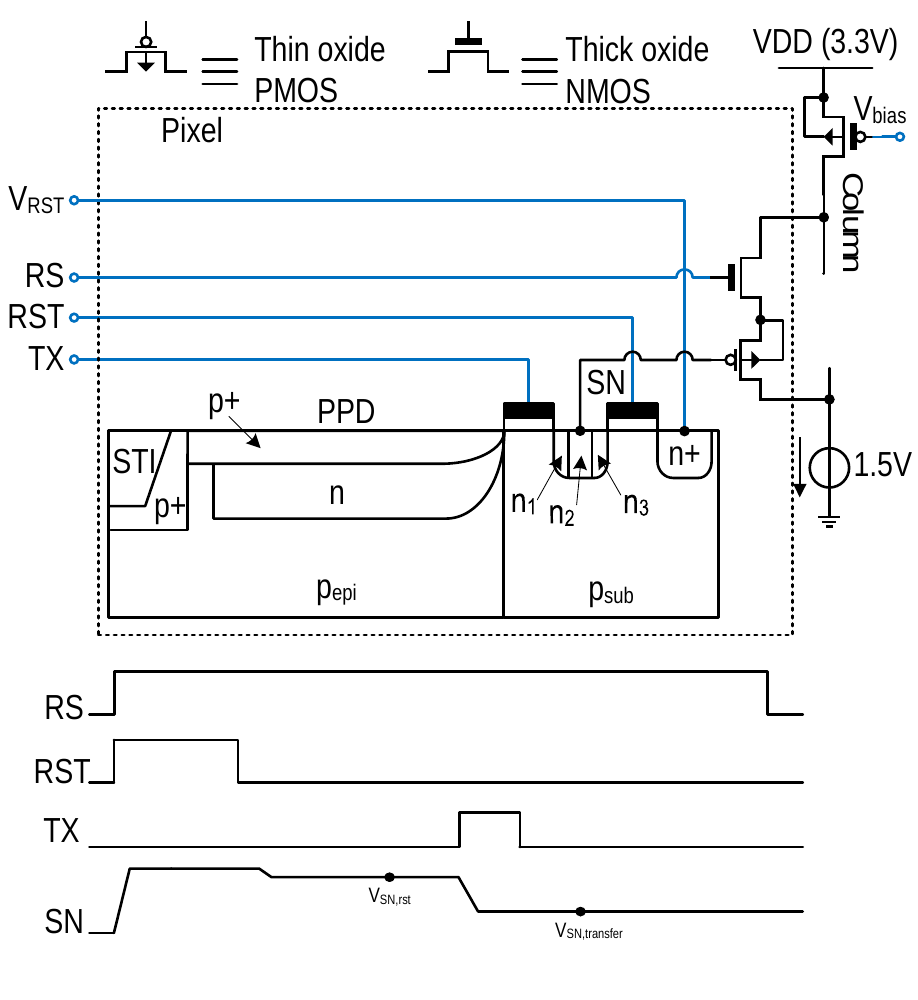}
\caption{schematic view and corresponding timing diagram of the second pixel variant featuring a 4T pixel scheme with a PMOS voltage follower having its source connected to its bulk for body effect mitigation.}\label{fig:4T_pixel_PMOS}
\end{figure}

\begin{figure}[]
\centering
\includegraphics[width=0.8\linewidth]{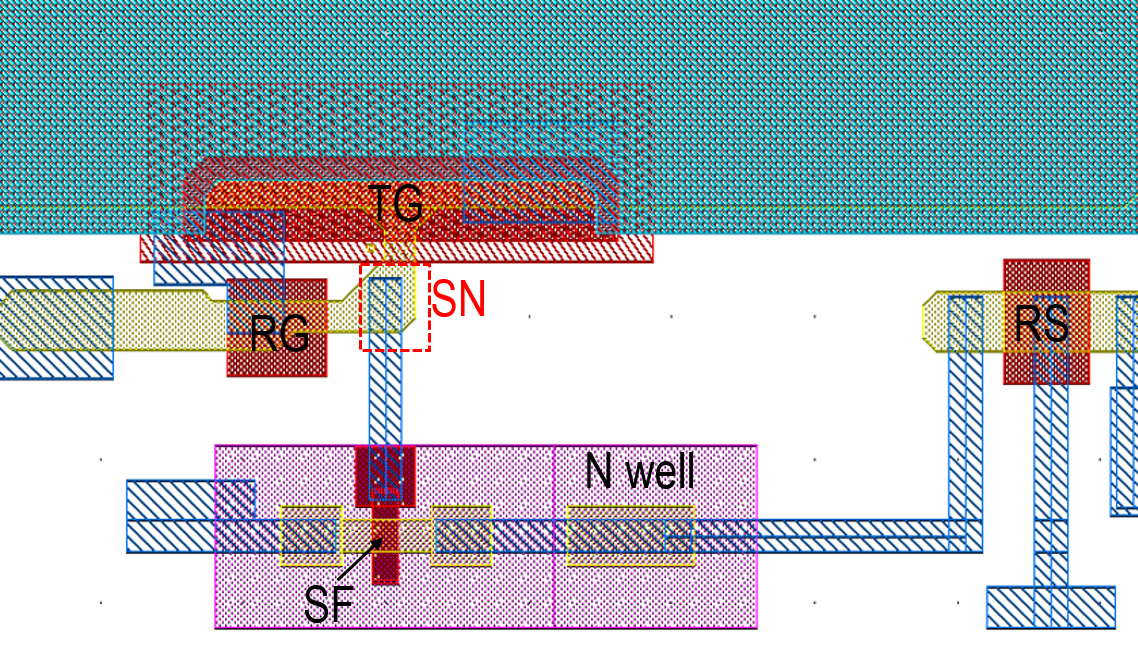}
\caption{layout view of the 4T pixel based on an PMOS SF and implementing the SN gradual doping.}\label{fig:pmos_layout}
\end{figure}

\subsection{Pixel variant 2}

After reducing the $C_{\text{SN}}$ term, the contribution of the SF parasitic capacitance to the $A_{\text{CG}}$ denominator is no more negligible Eq.\,\ref{eq:CG_SF}. Thus, the second $A_{\text{CG}}$ enhancement technique consists in optimizing the SF. The optimal SF sizing for a low input-referred 1/f and thermal noise is close to minimum sizing \cite{Bouk_NEWCAS_2014}. Due to the foundry design rules constraints, the NMOS SF size cannot be further reduced. Hence, a way to go around this limitation is to use thin oxide transistors that are available in the same design kit. Thin oxide transistors are 1.8\,V transistors featuring higher oxide density compared to the thick oxide ones used by default in pixel design. Even if these transistors feature higher oxide thickness, they allow to go for smaller gate width and length reducing consequently the parasitic capacitance. In order to use 1.8\,V transistors in a 3.3\,V design, the bulk voltage needs to be increased. PMOS transistors come with a private n-well with a bulk connection. By connecting the bulk to the source, the body effect is also mitigated which brings the slope factor $n$ in Eq.\,\ref{eq:CG_SF} close to 1 leading to a higher CG that can be approximated by:
\begin{equation}\label{eq:CG_SF_1}
A_{\text{CG}}=\frac{1}{C_{\text{SN}}+C_{\text{e}}W}.
\end{equation}
In this way using a thin oxide PMOS SF has both the advantage of featuring lower parasitic capacitance and body effect. The lower body effect increases the $A_{\text{CG}}$ by reducing $n$ and mitigating the gate-to-source total parasitic capacitance.\\
Fig.\,\ref{fig:4T_pixel_PMOS} shows the schematic and timing diagram of the pixel implementing a thin oxide PMOS SF. As for the previous suggested optimization, this pixel scheme does not have any impact on the timing diagram but rather requires and additional voltage reference connection shifting-up the SF drain to 1.5\,V in order to accommodate the 1.8\,V transistor to the 3.3\,V environment. On the layout side, the introduction of a private n-well for the PMOS SF faces additional design rules constraints reducing the pixel fill factor. Indeed a minimum spacing needs to be respected between the PPD well and the PMOS n-well as shown in Fig.\,\ref{fig:pmos_layout}. On the other hand the SF gate width and length can be reduced to a value as low as 0.2\,$\mu$m. In this pixel, a thinner metal wiring between the SN and the SF gate is also used in order to reduce the wiring parasitic capacitance with respect to the pixel variant previously introduced. With respect to the previous pixel variant, the fill factor is reduced to 61\% due to the additional distance to keep between the PMOS SF n-well and the PPD n-well.

\subsection{Pixel variant 3}

\begin{figure}[]
\centering
\includegraphics[width=0.8\linewidth]{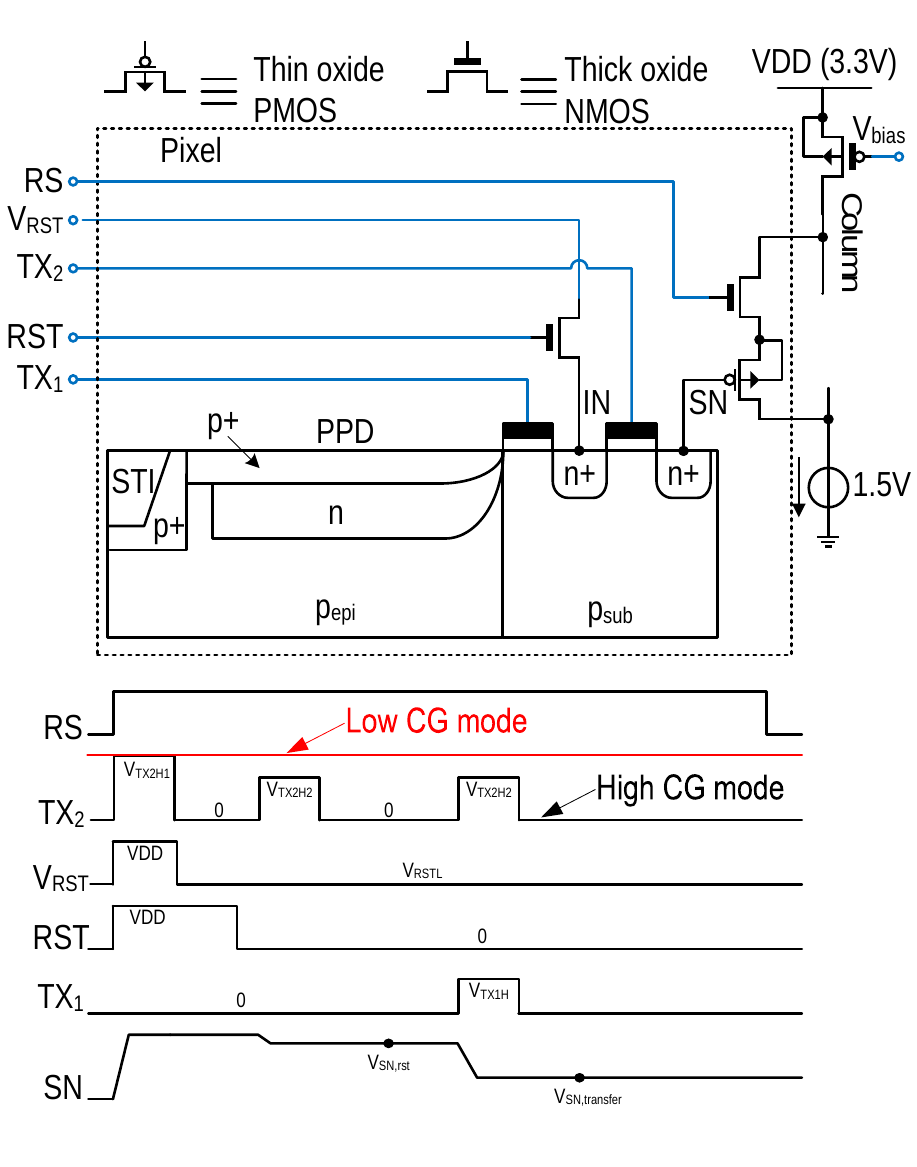}
\caption{Schematic view of the third pixel variant and corresponding timing diagram. The pixel features an unconventional 5T scheme completely isolating the SN from transfer and reset gates overlaps.}\label{fig:5T_pixel}
\end{figure}

\begin{figure*}[]
  \centering
  \subfigure[]
  {\label{fig:a}
  \includegraphics[width=0.4\linewidth]{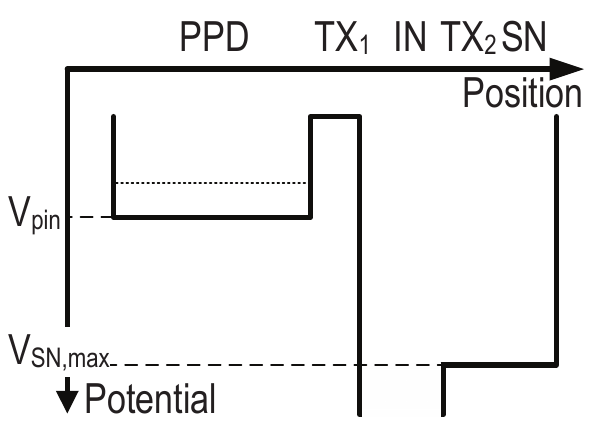}}
  \
  \subfigure[]
  {\label{fig:b}
  \includegraphics[width=0.4\linewidth]{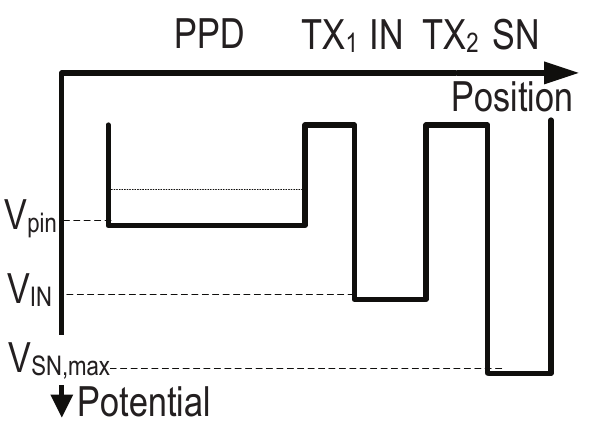}}
  \\
  \subfigure[]
  {\label{fig:c}
  \includegraphics[width=0.4\linewidth]{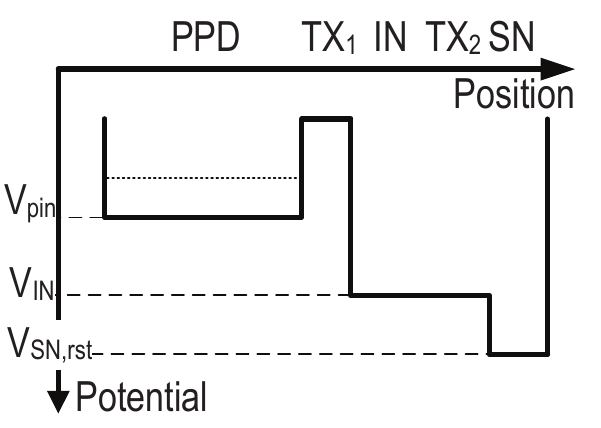}}
  \
    \subfigure[]
  {\label{fig:d}
  \includegraphics[width=0.4\linewidth]{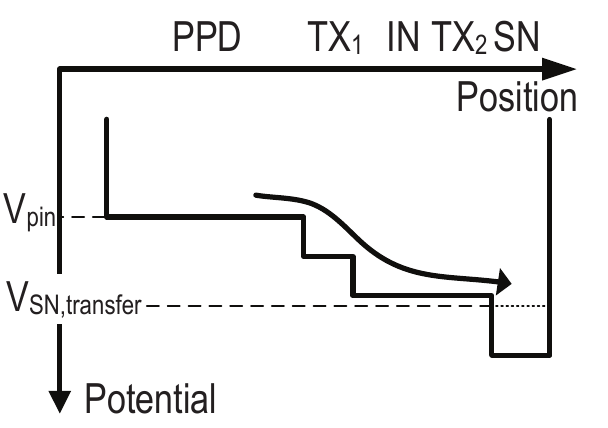}}

\caption{Idealistic potential profile across the photoelectrons path starting from the PPD, going to the IN through the transfer gate TX1 and then to the SN through the second gate TX2. (a) corresponds to the SN charge dump. (b) depicts the IN node setting to an intermediate voltage between the pin voltage and the SN one. (c) represents the SN reset and (d) the charge transfer from the PPD to the SN}\label{fig:Operation}
\end{figure*}

\begin{figure}[]
\centering
\includegraphics[width=0.8\linewidth]{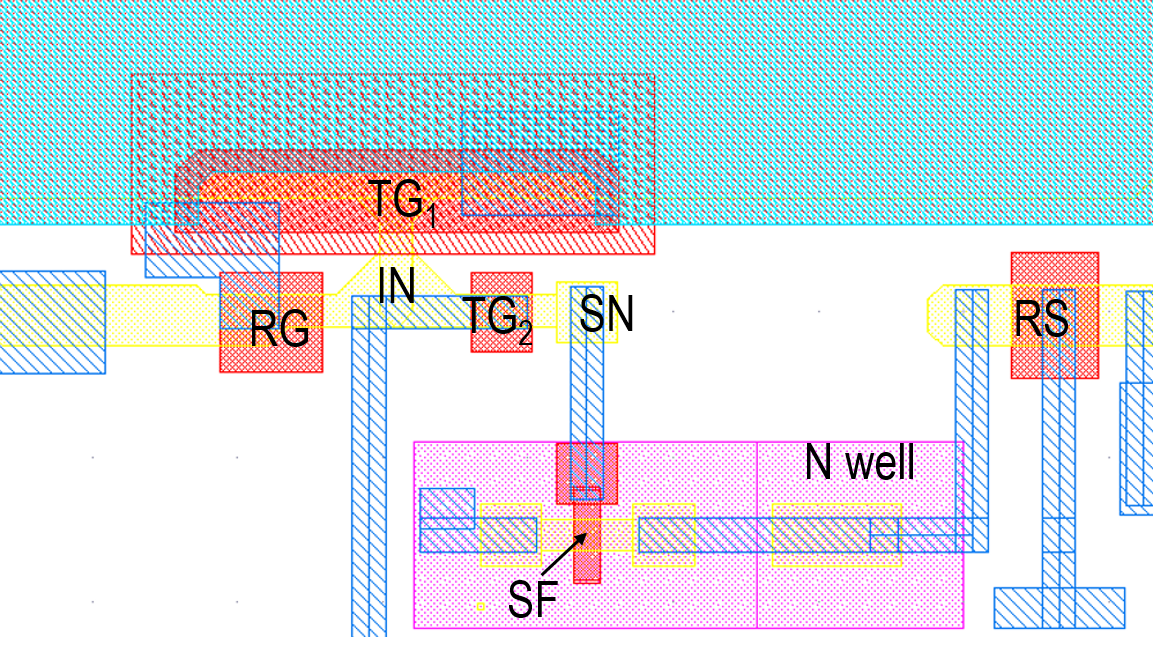}
\caption{layout view of the 4T pixel based on an PMOS SF and implementing the SN gradual doping.}\label{fig:pmos5T_layout}
\end{figure}

The gradual doping technique introduced in the previous pixels requires process refinements involving at least two additional implants with respect to a standard pixel. This represents a complexity barrier for the designers in addition to the manufacturing cost increase. Indeed, such refinement cannot by easily simulated \cite{Capp_TED_2019} as it is case for standard devices for which circuit simulators are used. An alternative technique to reduce the SN parasitic capacitance is to completely isolate it from the area overlapping with the transfer and reset gates by introducing an additional gate implemented in a standard process without involving any process refinements or additional implants.\\
Fig.\,\ref{fig:5T_pixel} shows a schematic view of the proposed 5T pixel and its corresponding timing diagram. Compared to a conventional 4T scheme, this pixel features an intermediate node (IN) that can be isolated from the SN by means of an additional gate allowing an important reduction of the SN capacitance. The reset phase consists in three steps. First, the RST switch is closed connecting IN to $\text{V}_{\text{RST}}$. While $\text{V}_{\text{RST}}$ is set to VDD, the potential barrier between IN and SN is lowered by setting $\text{TX}_{\text{2}}$ to a voltage $\text{V}_{\text{TX2H1}}$ in order to dump the charge from the SN as depicted in Fig.\,\ref{fig:a}. $\text{TX}_{\text{2}}$ is set back to 0 in order to split the IN and SN and freeze the SN voltage at its maximum level. $\text{V}_{\text{RST}}$ is then switched to a lower voltage $\text{V}_{\text{RSTL}}$ between the pin voltage of the PPD $\text{V}_{\text{pin}}$ and $\text{V}_{\text{SN,max}}$. After this step, the reset switch is opened again to freeze the IN voltage at a value $\text{V}_{\text{IN}}$ as depicted in Fig.\,\ref{fig:b}. The last step of the reset phase consists in setting $\text{TX}_{\text{2}}$ to a voltage $\text{V}_{\text{TX2H2}}$ making the barrier between the IN and SN equal or slightly higher than $\text{V}_{\text{IN}}$ as shown in Fig.\,\ref{fig:c}. In this way, any excess charge transferred to IN would diffuse towards the SN. After lowering back $\text{TX}_{\text{2}}$, the SN reset voltage $\text{V}_{\text{SN,rst}}$ is sensed. Transferring the charge integrated in the PPD to the SN takes place by pulsing both $\text{TX}_{\text{1}}$ and $\text{TX}_{\text{2}}$ as depicted in Fig.\,\ref{fig:d}. $\text{TX}_{\text{1}}$ is pulsed to a value $\text{V}_{\text{TX1H}}$ in order to set the voltage under the TG between the PPD pin voltage $\text{V}_{\text{pin}}$ and the intermediate node voltage $\text{V}_{\text{IN}}$ while $\text{TX}_{\text{2}}$ is pulsed again to transfer this charge to the SN. The signal corresponds to the difference between the SN voltage after reset $\text{V}_{\text{SN,rst}}$ and the one sensed after the transfer $\text{V}_{\text{SN,transfer}}$. A deeper description of the operation of this pixel is provided in \cite{Bouk_EDL_2020}. Fig.\,\ref{fig:pmos5T_layout} shows the layout of the proposed 5T pixel. With respect to the previous pixel, the additional transfer gate has no impact on the fill factor given the minimum spacing constraint related to the PMOS n-well.

\section{Measurements}

\begin{figure}[]
\centering
\includegraphics[width=0.8\linewidth]{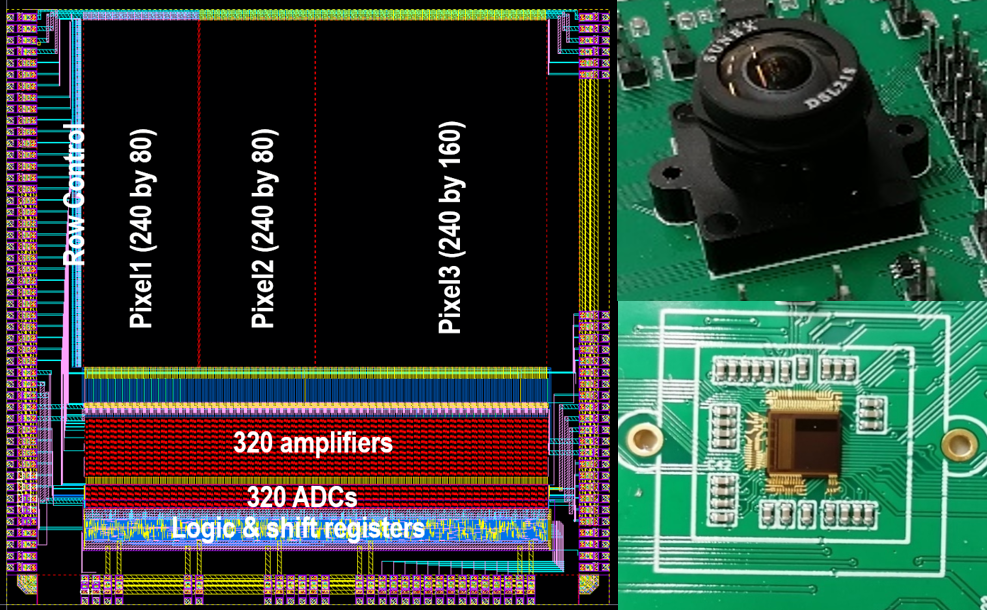}
\caption{Fabricated QVGA imager used for the measurements implementing the three introduced pixel variants. The image on the left shows the floor plan of the imager. The right images show the chip wire-bonded to the test PCB.}\label{fig:chip}
\end{figure}

\subsection{Physical implementation}
The three pixel variants, presented previously, are embedded in three sub-arrays of an image sensor chip fabricated in a 180\,nm CIS process with 4\,metal layers. The sensor chip measures 5\,mm by 5\,mm. Fig.\,\ref{fig:chip} shows the floor plan of the imager embedding a QVGA array made of 3 sub arrays exploiting the pixels presented in the previous sections with a pitch of 12\,$\mu$m. The imager also embeds row control circuits, column-level amplification and single slope 12 bits ADCs.\\
The imager chip is directly wire-bonded to the test printed circuit board (PCB) on which an optical objective is directly mounted on a fixed barrel as shown in Fig.\,\ref{fig:chip}. In order to perform pixel characterization, the optical objective is replaced by a light source mounted on a diffuser to make sure all pixels are similarly exposed.

\subsection{Measurement technique: photon transfer curve}
In order to measure the conversion gain, the photon transfer curve (PTC) method \cite{Jane_SPIE_2007} is used. This method exploits the proportionality between the shot noise variance and the average signal. For an average number of $N$ integrated photons, the variance of the corresponding shot noise is $N$.\\
The mean value of the signal at the output of the the pixel is given by
\begin{equation}\label{eq:Photon_Shot_Noise_Average}
E[V_{\text{out}}]= A_{\text{CG}} \cdot N.
\end{equation}
On the other hand, the variance of the output voltage when the photon shot noise dominates is given by
\begin{equation}\label{eq:Photon_Shot_Noise_Variance}
Var[V_{\text{out}}]= A_{\text{CG}}^2 \cdot N.
\end{equation}
Thus, the readout chain conversion gain can be obtained without knowing the exact value of $N$ combining Eq.\,\ref{eq:Photon_Shot_Noise_Average} and Eq.\,\ref{eq:Photon_Shot_Noise_Variance}
\begin{equation}\label{eq:CG_Measurement}
A_{\text{CG}} = \frac{Var[V_{\text{out}}]}{ E[V_{\text{out}}]}.
\end{equation}
Therefore, the pixel output variance plot as a function of the mean must feature a linear trend if the readout chain is shot noise limited. In that case, the slope of the linear trend corresponds to the conversion gain.\\
This technique is used to prove the shot noise limited performance obtained with all the pixels presented in this work and at the same time gives the evaluation of each pixel conversion gain. In order to obtain the PTC curve, each pixel is illuminated using a voltage controlled light emitting diode (LED) to obtain different points. For each LED voltage value, a 100 readouts are operated in order to compute the corresponding variance and mean values.

\subsection{Measurement results}

\begin{figure}[]
\centering
\includegraphics[width=0.8\linewidth]{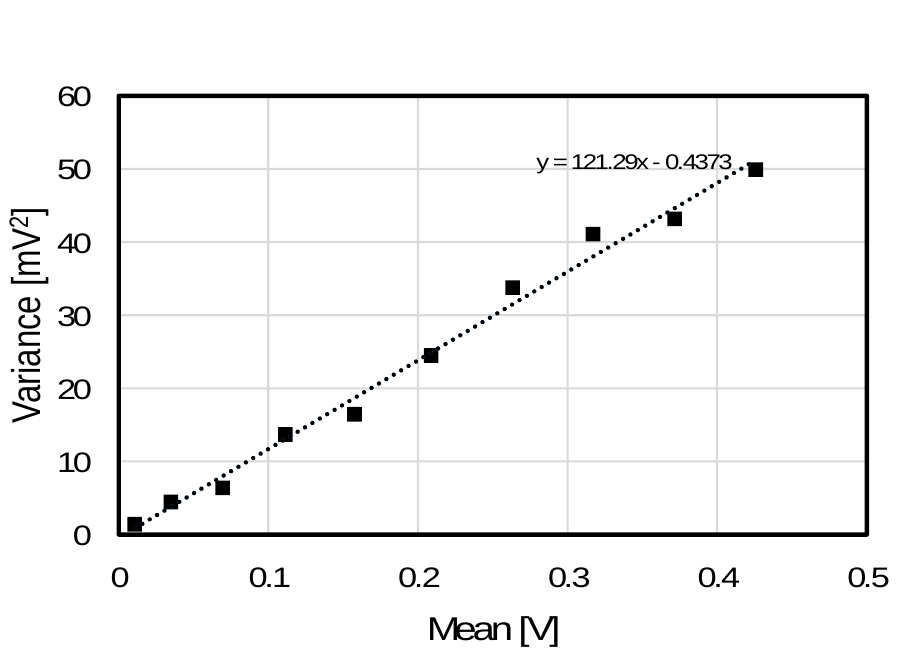}
\caption{PTC curve measured from the first variant pixel implementing NMOS SF and optimized SN doping.}\label{fig:PTC_4T_nmos}
\end{figure}

Fig.\,\ref{fig:PTC_4T_nmos} shows the PTC curve obtained from the 4T pixel implementing a NMOS SF and optimized SN doping. The PTC curve shows good linearity which demonstrates a low readout noise dominated by the shot noise even at low light conditions. The measured $A_{\text{CG}}$ corresponds to 121\,$\mu$V/e$^-$. A conventional 4T pixel based on the same SF but featuring a standard SN junction features a conversion gain of about 80\,$\mu$V/e$^-$ \cite{Capp_TCAS_2020,Bouk_TED_2015}.\\
Going back to the $A_{\text{CG}}$ formula introduced in Eq.\,\ref{eq:CG_SF}, the slope factor of the NMOS SF is about 1.2 based on the simulation results. The SF gate intrinsic capacitance can be approximated by 2/3C$_{\text{ox}}$WL, hence, for a C$_{\text{ox}}$ of 4.5\,fF/$\mu$m$^\text{2}$ a width of 0.4\,2$\mu$m and a length of 0.84\,$\mu$m the source follower gate capacitance contribution to $A_{\text{CG}}$ denominator is 0.2\,fF. The extrinsic capacitance contribution dominated by the overlap between the source and drain of the SF is also estimated from the simulation to be as high as 0.2\,fF. The measured total denominator capacitance corresponding to the 121\,$\mu$V/e$^-$ conversion gain corresponds to 1.1\,fF. Hence the $\text{C}_{\text{SN}}$ term corresponds to 0.7\,fF. This same term is as high as 1.2\,fF in the standard pixel which means that the gradual doping technique used at the level of the SN reduces $\text{C}_{\text{SN}}$ by a factor 1.7.

\begin{figure}[]
\centering
\includegraphics[width=0.8\linewidth]{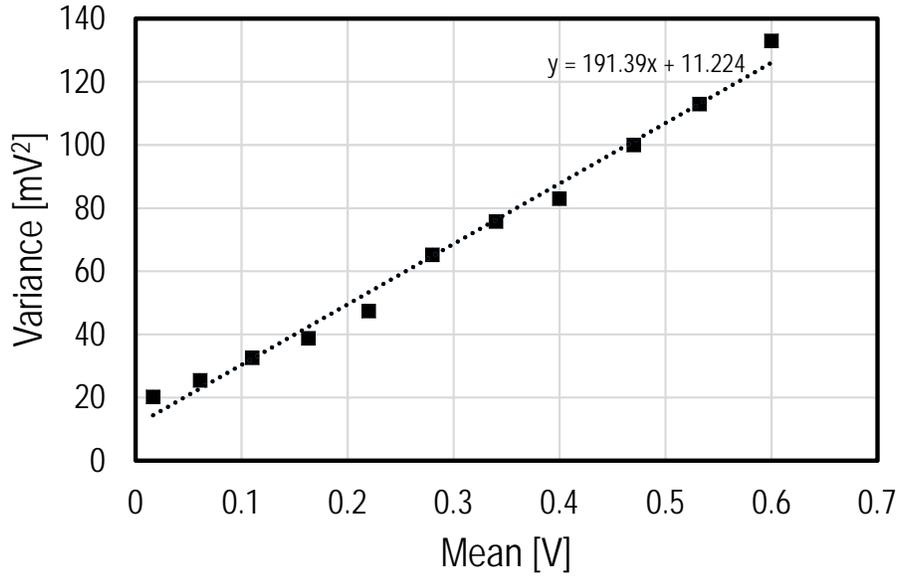}
\caption{PTC curve measured from the second pixel variant implementing PMOS SF and optimized SN doping.}\label{fig:PTC_4T_pmos}
\end{figure}

Fig.\,\ref{fig:PTC_4T_pmos} shows the PTC curve obtained from the 4T pixel implementing the same SN structure as the previous pixel and replacing the NMOS SF by an thin oxide PMOS SF with a source-to-bulk connection. Connecting the source to bulk makes the slope factor approximately equal to 1 which increases the conversion gain. The PTC curve shows good linearity which demonstrates a low readout noise dominated by the shot noise even at low light conditions. The measured conversion gain corresponds to 191\,$\mu$V/e$^-$. In this case, the contribution of the SF to the conversion gain denominator is reduced to 0.2\,fF given that the (1-1/n) term is close to 0, $\text{C}_{\text{i}}$ equal to 0.8\,fF and the width of the SF is equal to 0.22\,$\mu$m. Hence the $\text{C}_{\text{SN}}$ term is estimated to 0.6\,fF, which is quite close to the 0.7\,fF obtained from the previous pixel. In fact, both pixels share the same SN structure with a slight difference on the metal line connecting the SN to the SF as shown in the layouts of both pixels.

\begin{figure}[]
\centering
\includegraphics[width=0.8\linewidth]{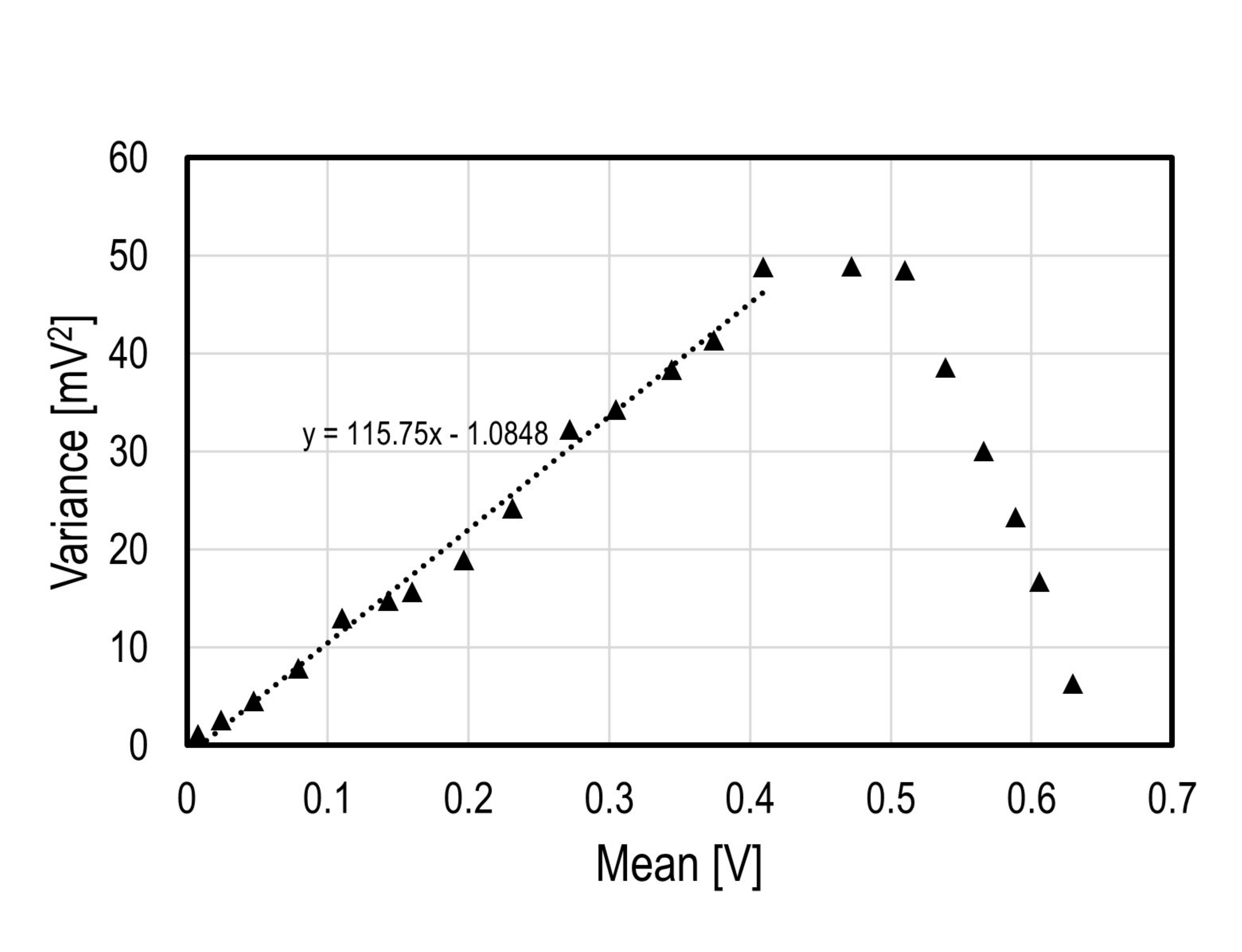}
\caption{PTC curves measured from the 5T pixel in low conversion gain mode when the SN and IN are merged.}\label{fig:PTC_5T_LG}
\end{figure}

\begin{figure}[]
\centering
\includegraphics[width=0.9\linewidth]{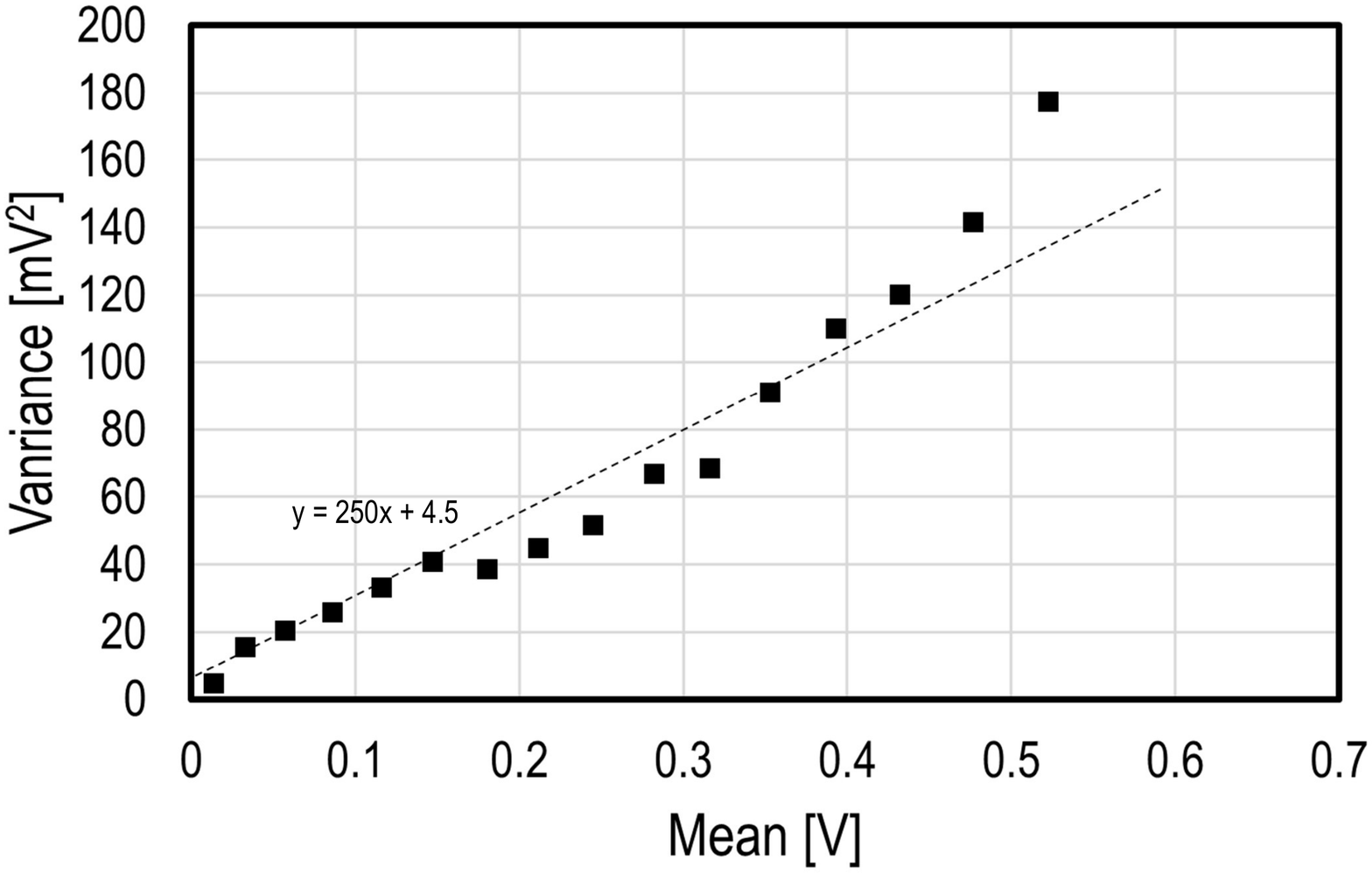}
\caption{PTC curves measured from the 5T pixel in high conversion gain mode, completely isolating the SN from transfer and reset gates overlap.}\label{fig:PTC_5T_HG}
\end{figure}

\begin{figure}[]
\centering
\includegraphics[width=0.5\linewidth]{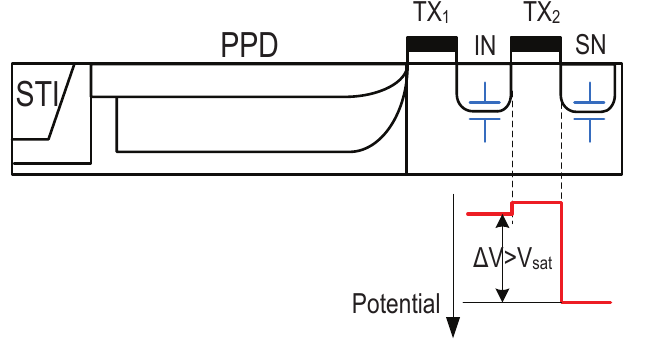}
\caption{Simplified model of the IN, SN and intermediate TG depicting the corresponding potential levels during charge transfer.}\label{fig:TX2}
\end{figure}

For the 5T pixel, Fig.\,\ref{fig:PTC_5T_LG} shows the measured PTC for the low gain mode. In this configuration, the IN and SN are completely merged. The PTC curve show a linear response which demonstrates a low readout noise dominated by the shot noise even at low light conditions. The linear region of the curve points at measured conversion gain of 115\,$\mu$V/e$^-$. This PTC collapses around 0.65\,V which corresponds to a FWC of about 5600 electrons. Note that this FWC can be increased by increasing the IN capacitance.

Fig.\,\ref{fig:PTC_5T_HG} shows the measured PTC for the high gain mode following the timing diagram of Fig.\,\ref{fig:5T_pixel} and setting V$_\text{RSTL}$ to 1.76\,V, V$_\text{TX1H}$ to 1.2\,V, V$_\text{TX2H1}$ to 2.2\,V and V$_\text{TX2H2}$ to 1.5\,V.
In this configuration, the SN is completely isolated from the transfer and reset gates overlaps thanks to the second TG. As shown in Fig.\,\ref{fig:TX2}, the IN and SN can be modeled by capacitors connected to the source and drain of the second TG. Before the charge transfer, as explained previously, the voltage difference between the IN and SN is cosen to set to the second TG in saturation regime. As shown by Fig.\,\ref{fig:PTC_5T_HG}, the high CG curve follows a linear trend corresponding to a slope of 250\,$\mu$V/e$^-$. The PTC deviates from its linear trend for higher charge transfers suggesting that other noise mechanisms dominate when relatively large amount of charge is transferred to the IN. This phenomena can be explained as follows: as long as long as the charge transferred to the SN does not reduce the SN voltage enough to become close to the IN voltage, the second TG operates in saturation regime. In this case, the second TG enters in a sub-threshold  reducing the kTC noise to a negligible level and giving only rise to the shot noise. In contrast, when a larger amount of charge is transferred, the IN and SN are merged and the second TG operates in an triode regime. This situation gives also rise to  spill back noise when closing the second TG. The high gain mode is expected to be used only for low light conditions in which the PTC remains linear. 

In the case of high gain mode, a conversion gain of 250\,$\mu$V/e$^-$ with the same SF as the previous pixel corresponds to a SN capacitance $C_{\text{SN}}$ of 0.44\,fF.

\subsection{Summary and discussion}

\begin{figure}[]
\centering
\includegraphics[width=1\linewidth]{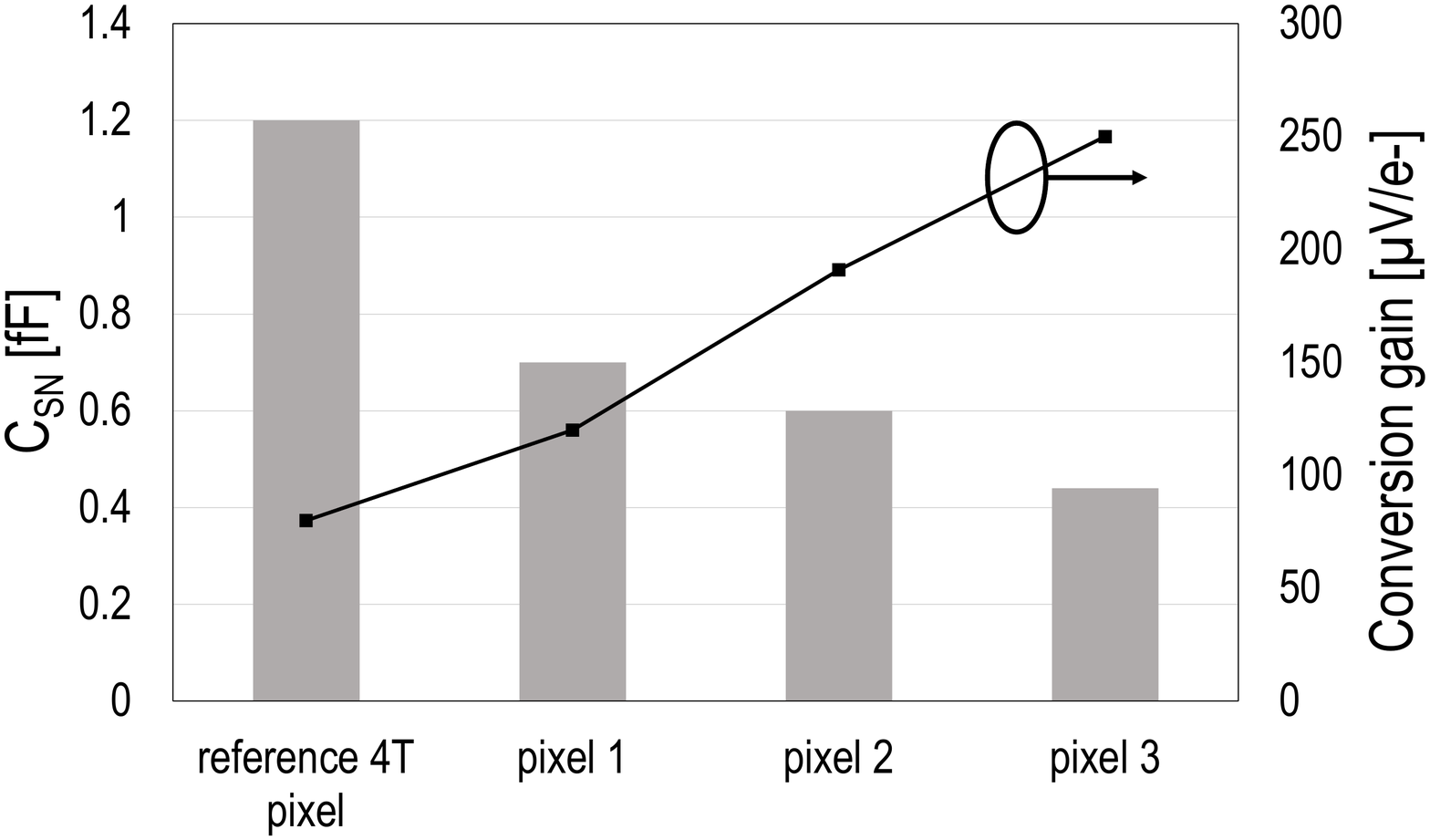}
\caption{Summary of the measured conversion gains $A_{\text{CG}}$ (right axis) and extracted sense node capacitance $C_{\text{SN}}$ (left axis) for the three pixel variants implemented in this work compared with a reference standard 4T pixel from the same foundry.}\label{fig:summary}
\end{figure}

The enhancement of the conversion gain relies both on the SN capacitance reduction and the SF optimization. Tab.\,\ref{tab:CG_evolution} summarizes the techniques deployed to increase the conversion gain and their impact. The measurement show a dramatic gradual increase by a factor of more than 3 going from the standard pixel to the 5T pixel scheme.

On the other hand, in a carefully designed CIS readout chain, the $1/f$ and thermal noise originating from the SF are the dominating noise sources \cite{Bouk_TED_2015,Bouk_Springer_2018}. The parameter that is more important for reducing the noise originating from the SF stage is the sense node parasitic capacitance independent from the SF contribution $C_{\text{SN}}$. Indeed, based on Eq.\,\ref{eq:Input_Ref_1/fNoise} and Eq.\,\ref{eq:Input_Ref_thermal_Noise}, if the SF is minimally sized, the $C_{\text{SN}}$ term dominates the SF parasitic capacitance contribution and its reduction directly mitigates the SF stage noise contribution. Tab.\,\ref{tab:SN_evolution} summarizes the different techniques implemented to reduce the sense node capacitance and their impact on $C_{\text{SN}}$. The $C_{\text{SN}}$ values extracted from the measurement show a progressive decrease of the SN parasitic capacitance.

Fig.\,\ref{fig:summary} summarizes the measurement and extraction results. The measured CG increase with respect to a reference standard 4T pixel comes with a reduction of the $C_{\text{SN}}$.

\begin{table}[]
\caption{Conversion gain evolution with the combination of enhancement techniques}
\centering
\begin{tabular}{p{0.7\linewidth} p{0.2\linewidth}  }
\toprule
 Implemented technique & $A_{\text{CG}}$ [$\mu$V/e$^-$]   \\

\midrule

standard 4T pixel fabricated with the 180\,nm CIS process used in this work &  80  \\
\midrule
gradual sense node doping and standard NMOS SF & 120 \\
\midrule
gradual sense node doping combined with PMOS SF with source connected to the bulk for body effect mitigation  & 191   \\
\midrule
5T pixel scheme with isolated SN and PMOS SF with mitigated body effect   & 250 \\

\bottomrule
\end{tabular}
\label{tab:CG_evolution}
\end{table}
\begin{table}[]
\caption{Sense node capacitance evolution with different reduction techniques.}
\centering
\begin{tabular}{p{0.7\linewidth} p{0.2\linewidth}  }
\toprule
 Implemented technique & $C_{\text{SN}}$ [fF]   \\

\midrule

standard 4T pixel fabricated with the 180\,nm CIS process used in this work &  1.2  \\
\midrule
gradual sense node doping reducing the junction and overlap capacitance with the transfer and reset gates & 0.6 to 0.7   \\
\midrule
isolation of the SN from the reset and transfer gates by means of an additional transfer gate in 5T pixel scheme  & 0.44   \\

\bottomrule
\end{tabular}

\label{tab:SN_evolution}
\end{table}


\section{Conclusion}

The pixel conversion gain is a key performance metric in low noise CIS readout chains. It is both affected by the SN total parasitic capacitance and the in-pixel SF design.\\
When metal wiring parasitic capacitance is low enough thanks to an optimal layout, the SN parasitic capacitance can efficiently be reduced by mitigating the impact of the overlapping capacitance with the transfer and reset gates connected to the SN. This is successfully achieved by implementing an optimized doping profile in the SN area. The latter consists of using light doping in general combined with gradual doping concentrations having an effect similar to LLD for MOS transistors. Such a technique brings the SN parasitic capacitance from 1.2\,fF in a standard pixel to 0.6\,fF.\\
this technique requires process refinements coming with complex research and development  effort since they are not straightforward to simulate. An alternative to this technique is presented. It involves the complete isolation of the SN from the transfer and reset gates by means of an additional gate. This technique uses the standard process and successfully brings the SN capacitance to 0.44\,fF at the cost of a higher lag.\\
On the SF design side, the mitigation of the body effect successfully increases the overall CG. Indeed, the CG goes from 120\,$\mu$V/e- to 191\,$\mu$V/e$^-$ when replacing the standard NMOS SF by a source to bulk connected PMOS in the pixel embedding the gradual SN doping technique.\\
Finally, combining both SN isolation from the TG with an optimized SF leads to a CG of 250\,$\mu$V/e$^-$ which is 3 times higher than a standard 4T pixel from the same process.


\bibliography{bibliography_CIS}
\bibliographystyle{spiejour}


\vspace{2ex}\noindent\textbf{Assim Boukhayma} is chief scientifc officer at Senbiosys, Switzerland. He is also guest scientist and academic advisor for EPFL, Switzerland. He received the graduate engineering degree in telecommunications and  M.Sc. degree, in microelectronics and embedded systems architecture, from Institut Mines Telecom, Brest, France in 2013. He received the Ph.D. degree
from the Ecole Polytechnique Federale de Lausanne (EPFL), Lausanne, Switzerland, on the subject of Ultra Low Noise CMOS Image Sensors in 2016. In 2018, he co-founded Senbiosys, Neuchatel, Switzerland.\\
From 2012 to 2016, he was a scientist at Commissariat a l'Energie Atomique (CEA-LETI),Grenoble, France, in the frame of his PhD research. In 2012, he did his M.Sc. Internship at
CEA-LETI on the design of a low-noise CMOS THz camera.\\
Assim's awards and honors include the french DGA scholarship, the Springer thesis award for outstanding PhD research and  the IEEE sensors journal best paper award runner up.

\vspace{1ex}
\noindent Biographies and photographs of the other authors are not available.

\listoffigures
\listoftables

\end{spacing}
\end{document}